\documentclass[prd,nofootinbib,
, twocolumn
, nofootinbib
, superscriptaddress
]{revtex4-2}

\usepackage{aas_macros}
\usepackage{graphicx}
\usepackage[caption=false]{subfig}
\usepackage{siunitx}
\usepackage{mathrsfs}
\usepackage{amsmath,amssymb}
\usepackage{bm}
\usepackage{braket}
\usepackage{listings}
\usepackage{cases}
\usepackage{here}
\usepackage{comment}
\usepackage{soul}
\usepackage{cancel}
\usepackage{cases}
\usepackage[utf8]{inputenc}
\usepackage{url}
\usepackage{longtable}
\usepackage[normalem]{ulem}
\usepackage{xspace}
\usepackage[colorlinks=true
,urlcolor=blue
,anchorcolor=blue
,citecolor=blue
,filecolor=blue
,linkcolor=blue
,menucolor=blue
,linktocpage=true
,pdfproducer=medialab
,pdfa=true
]{hyperref}
\usepackage{animate}
\newcommand{\eq}[1]{\begin{equation}\begin{split} #1 \end{split}\end{equation}}

\newcommand{\lr}[1]{\left( #1 \right)}

\newcommand{\Ms}{M_{\odot}}

\newcommand{\mr}{\mathrm}

\newcommand{\as}[1]{\SI{ #1 }{\arcsecond}}

\begin{document}
%\title{Fluctuations of macro critical curves by subhalos}
\title{Analytic approach to astrometric perturbations of critical curves by substructures}
\author{Katsuya T. Abe}
\email{kabe@chiba-u.jp}
\affiliation{Center for Frontier Science, Chiba University, 1-33 Yayoi-cho, Inage-ku, Chiba 263-8522, Japan}
\author{Hiroki Kawai}
%\email{}
\affiliation{Department of Physics, Graduate School of Science, Chiba University, 1-33 Yayoi-Cho, Inage-Ku, Chiba 263-8522, Japan}
\affiliation{Department of Physics, School of Science, The University of Tokyo, Bunkyo, Tokyo 113-0033, Japan}
\affiliation{INAF – Osservatorio Astronomico di Bologna, via Ranzani 1, 40127 Bologna, Italy}
\author{Masamune Oguri}
%\email{}
\affiliation{Center for Frontier Science, Chiba University, 1-33 Yayoi-cho, Inage-ku, Chiba 263-8522, Japan}
\affiliation{Department of Physics, Graduate School of Science, Chiba University, 1-33 Yayoi-Cho, Inage-Ku, Chiba 263-8522, Japan}

\begin{abstract}
Astrometric perturbations of critical curves in strong lens systems are thought to be one of the most promising probes of substructures down to small-mass scales. 
While a smooth mass distribution creates a symmetric geometry of critical curves with radii of curvature about the Einstein radius, substructures introduce small-scale distortions on critical curves, which can break the symmetry of gravitational lensing events near critical curves, such as highly magnified individual stars.
We derive a general formula that connects the fluctuation of critical curves with the fluctuation of the surface density caused by substructures, which is useful when constraining models of substructures from observed astrometric perturbations of critical curves. We numerically check that the formula is valid and accurate as long as substructures are not dominated by a small number of massive structures.
As a demonstration of the formula, we also explore the possibility that an anomalous position of an extremely magnified star, recently reported as ``Mothra,'' can be explained by fluctuations in the critical curve due to substructures. 
We find that cold dark matter subhalos with masses ranging from $5 \times 10^7 \Ms/h$ to $10^9 \Ms/h$ can well explain the anomalous position of Mothra, while in the fuzzy dark matter model, the very small mass of $\sim 10^{-24}$~eV is needed to explain it.
\end{abstract}

\maketitle

\section{Introduction}

Substructures within dark matter~(DM) halos have been intensively studied in recent decades, motivated by a wide range of topics in cosmology.
The abundance and distribution of substructures provide valuable insight into primordial perturbations, which are the initial seeds of cosmological structures.
While observations of the cosmic microwave background and the large-scale structures have tightly constrained statistical properties of primordial perturbations on scales approximately larger than $1$~Mpc, or roughly higher than $10^{13} \Ms$ in terms of mass of DM halos, our understanding of smaller scales, specifically those corresponding to substructures, remains limited~\cite{2010MNRAS.404...60R,2019JCAP...07..017C, 2019MNRAS.489.2247C, 2020A&A...641A...6P}.
% Since primordial perturbations originated in the very early Universe, investigating substructures also provides a unique opportunity to gain deeper insights into the inflationary mechanisms~\cite{2016ASSP...45...41M}.

Properties of substructures can also potentially reflect the properties of DM.
While current observations, especially of large-scale structures, strongly support cold dark matter~(CDM), several deviations from the CDM prediction have been reported at smaller scales~(see Refs.~\cite{2017Galax...5...17D, 2017ARA&A..55..343B} for review), which may hint alternatives to CDM, such as warm dark matter~\cite{2000ApJ...542..622C, 2001ApJ...556...93B, 2005PhRvD..71f3534V} and dark matter with a macroscopic de Broglie wavelength, referred to as fuzzy dark matter~(FDM)~\cite{2000ApJ...534L.127P,2000PhRvL..85.1158H,PhysRevLett.113.261302, 2017PhRvD..95d3541H}. 
%The question of the nature of dark matter continues to be debated over the years.
In the CDM paradigm, masses of substructures span a wide range, from approximately $10^{-6} \Ms$ to around $10^{16} \Ms$. 
In contrast, warm dark matter or FDM predicts characteristic mass scales that preclude the formation of smaller substructures. 
FDM also predicts specific shapes of substructures, such as clumps resulting from quantum interference patterns and cores resulting from quantum pressure, often referred to as soliton cores. Therefore, detailed studies of substructures down to small mass scales may provide an important clue to the nature of DM.

One promising method to study substructures down to small mass scales is to use gravitational lensing effects~\cite{1999ApJ...524L..19M}.
In particular, strong gravitational lensing effects, which exhibit high magnification and multiple images of the distant source, are affected by substructures inside primary lens objects, such as massive galaxies and galaxy clusters.
Several studies pointed out a link between strong gravitational lensing systems with anomalous flux ratios and substructures inside lens objects~\cite{1998MNRAS.295..587M, 2001ApJ...563....9M, 2002ApJ...565...17C,  2002ApJ...567L...5M, 2002ApJ...572...25D, 2002A&A...388..373B, 2012MNRAS.419.3414M, 2017MNRAS.471.2224N, 2020MNRAS.491.6077G, 10.1093/mnras/stz3177}.
In Ref.~\cite{10.1093/mnras/stz3177}, they analyzed seven gravitational lensing events of radio quasars which show flux anomalies using the lens model, including substructures, stellar discs, and line-of-sight haloes.
Then they found a mass fraction of substructures $f_{\mr{sat}}$ as $0.8\% < f_{\mr{sat}} < 1.9\%$, at the $68\%$ confidence level, with a median estimation of $f_{\mr{sat}} = 1.2\%$, which is in agreement with the predictions from CDM hydrodynamical simulations within $1\sigma$.
% In Ref.~\cite{2002ApJ...572...25D}, they analyzed seven gravitational lensing events of radio quasars which show flux anomalies using the lens model, including substructures.
% Then they found a mass fraction of substructures $f_{\mr{sat}}$ as $0.6\% < f_{\mr{sat}} < 7\%$, at the $90\%$ confidence level, with a median estimation of $f_{\mr{sat}} = 2\%$.
In addition, substructures can also be probed by observing the distortions in the surface brightness patterns of lensed giant arcs~\cite{2013ApJ...767....9H, 2016JCAP...11..048H, 2017JCAP...05..037B, 2017MNRAS.472..129A}.

Reference~\cite{2018ApJ...867...24D} proposed a new method to search for substructures inside galaxy clusters focusing on astrometric perturbations.
A smooth mass distribution creates a geometry that yields multiple images with symmetric configurations and critical curves with radii of curvature about the Einstein radius. Hereafter, we refer to such lens models as ``macro''-lens models.
The existence of substructures introduces small-scale distortions on such a macrocritical curve, which can distort or break the symmetry of the gravitational lensing events expected from the macrocritical curve. Such distortions of macrocritical curves can be probed by highly magnified individual stars that are observed in the vicinity of critical curves~\cite{2018NatAs...2..334K,2022Natur.603..815W}.
%They conducted a case study against the giant arc at $z = 0.725$ behind the lensing cluster, ``Abell 370", at $z = 0.375$.
It is argued that CDM subhalos of masses ranging from $10^6\Ms$ to $10^8\Ms$ produce fluctuations of the macrocritical curve that could be detected with $\sim10$ hr integrations with the \textit{James Webb Space Telescope}~(JWST) in near-infrared bands (see also Ref.~\cite{2023arXiv230406064W}).

% In this paper, we derive a general formula that links fluctuations in the macrocritical curve to the surface density caused by the substructures, allowing us to analytically estimate the amplitude of the fluctuations from the surface density power spectrum of the substructures. 
In this paper, we derive a general formula that connects fluctuations in macrocritical curves with the fluctuation of the surface density caused by substructures. 
This formula allows us to analytically estimate the amplitude of the fluctuations from the surface density power spectrum of substructures, which is useful for discussing the mass of the substructures in the context of astrometric perturbations.
As for the surface density power spectrum, previous studies have proposed the expressions for the different types of substructures such as CDM subhalos~~\cite{2016JCAP...11..048H} and quantum clumps of FDM~\cite{2022ApJ...925...61K}. 
We verify the validity of this formula by performing numerical simulations. 

% Recently, \citet{2023arXiv230710363D} reported an extremely magnified (likely binary) star, formally named ``LS1" and also nicknamed ``Mothra", in the lensed arc of the galaxies behind the galaxy cluster MACS0416. The binary star shows an anomaly in which the counterimage is not seen yet even though nearby stars are found in set with their counterimages (see Fig.~\ref{fig: mothra}).
% They have also argued that the anomaly of LS1 can be explained by considering a local millilensing effect caused by a substructure at the LS1 point. 
% However, in this paper, we attempt to interpret it as caused by fluctuations in the macrocritical curve by substructures such as the CDM subhalos and quantum clumps of FDM.
Recently, \citet{2023arXiv230710363D} reported an anomaly of an extremely magnified (likely binary) star, nicknamed ``Mothra,'' in which the counterimage is not seen yet even though nearby stars are found in set with their counterimages (see Fig.~2 in Ref.~\cite{2023arXiv230710363D}). 
% Recently, \citet{2023arXiv230710363D} reported an anomaly of an extremely magnified (likely binary) star, nicknamed ``Mothra," in which the counterimage is not seen yet even though nearby stars are found in set with their counterimages (see Fig.~\ref{fig: mothra}). 
They have argued that the anomaly could be explained by considering a local millilensing effect caused by a substructure. In this paper, as an immediate application of our formula, we discuss the possibility of explaining the Mothra's anomalous observed position by fluctuations in the macrocritical curve due to substructures such as CDM subhalos and quantum clumps of FDM.

This paper is organized as follows.
In Sec.~\ref{sec: fluct_cc}, we first review the macrolens model and derive the general formula connecting fluctuations in macrocritical curves with the fluctuation of the surface density caused by substructures from the lens equation. In Sec.~\ref{sec: validity_cc}, we numerically verify the formula and examine a parameter region in which this formula is accurate. 
In Sec.~\ref{sec: mothra}, we attempt an alternative interpretation of the anomalous lensing events utilizing the formula. In Sec.~\ref{sec: conclude}, we conclude this paper.

Throughout this paper, we assume a flat $\Lambda$CDM cosmology and fix cosmological parameters to the Planck 2018 best-fit values~\cite{2020A&A...641A...6P}.
% , ($h=0.6727$, $\omega_{\mathrm{b}}=0.02236, \omega_{\mathrm{cdm}}=0.1202, \ln \left(10^{10} A_{\mathrm{s}}\right)=3.045$, $n_{\mathrm{s}}=0.9649$, and $\tau_{\text {reio}}=0.0544$)

\section{Fluctuations of macrocritical curves}\label{sec: fluct_cc}
To study fluctuations of macrocritical curves by substructures, let us first define a macrolens model in which we set the origin of the coordinate system in the image and source planes on the critical curve and the caustic, respectively.

The lens equation relates the source position $\bm{\beta}=(\beta_x,\beta_y)$ with the image position $\bm{\theta}=(\theta_x,\theta_y)$ as
\eq{
\bm{\beta} = \bm{\theta} -\nabla\psi(\bm{\theta}),
}
with $\psi(\bm{\theta})$ being the lens potential.  By expanding the lens potential up to the third order at the origin, we obtain
\eq{
\begin{aligned}
\psi(\boldsymbol{\theta}) & =\frac{1}{2}\left(\psi_{,xx} \theta_{x}^2+2 \psi_{,xy} \theta_{x} \theta_{y}+\psi_{,yy} \theta_{y}^2\right) \\
& +\frac{1}{6}\left(\psi_{,xxx} \theta_{x}^3+3 \psi_{,xxy} \theta_{x}^2 \theta_{y}+3 \psi_{,xyy} \theta_{x} \theta_{y}^2+\psi_{,yyy} \theta_{y}^3\right),
\end{aligned}
}
where the subscripts $,x$ and $,y$ represent the derivative with respect to $\theta_x$ and $\theta_y$, respectively. Here, we drop the leading two terms by setting the origin of the image plane to the origin of the source plane. 
We denote the convergence at the origin as $\kappa_0$, which satisfies
\eq{\label{eq: phi11_plus_phi22}
\kappa_0 = \frac{1}{2}(\psi_{,xx}+\psi_{,yy}).
}
With the condition that the critical curve and the caustic pass through the origin and using Eq.~\eqref{eq: phi11_plus_phi22}, one can obtain
\eq{
& \psi_{,xx}=\kappa_0+\left(1-\kappa_0\right) \cos \omega, \\
& \psi_{,yy}=\kappa_0-\left(1-\kappa_0\right) \cos \omega, \\
& \psi_{,xy}=-\left(1-\kappa_0\right) \sin \omega,
}
using the arbitrary constant parameter $\omega$.
For simplicity, we set $\omega=0$ in this work. Additionally, we henceforward consider a complete orthogonal coordinate system in which the critical curve and multiple images are perpendicular to each other by setting $\psi_{,xxy} = \psi_{,xyy} = \psi_{,yyy} = 0$. 
Denoting $\psi_{,xxx} = \epsilon$, the lens potential can be written by
\eq{\label{eq: deflection_potential}
\psi(\boldsymbol{\theta})=\frac{1}{2}\left[\kappa_0\left(\theta_x^2+\theta_y^2\right)+\left(1-\kappa_0\right)\left(\theta_x^2-\theta_y^2\right)\right]-\frac{\epsilon}{6} \theta_x^3.
}
With this lens potential, the lens equation becomes very simple as
\eq{
\begin{aligned}
& \beta_x=\frac{\theta_{x}^2}{2}\epsilon, \\
& \beta_y=2\left(1-\kappa_0\right) \theta_{y}.
\end{aligned}
}
Note that $\epsilon$ has the dimension of the inverse of the angle, whose value approximately corresponds to the inverse of the Einstein radius of the macrolens model.
The Jacobian matrix is written by
\eq{\label{eq: Jmatrix_0th}
\lr{\frac{\partial \bm{\beta}(\bm{\theta})}{\partial \bm{\theta}}}=
\begin{pmatrix}
\epsilon \theta_{x} & 0 \\
0 & 2(1-\kappa_0) \\
\end{pmatrix}.
}
One can find that the $\theta_{y}$axis corresponds to the critical curve by calculating the determinant of the matrix in Eq.~\eqref{eq: Jmatrix_0th}.

Now, let us consider the fluctuations of the macrocritical curve due to substructures.
Considering fluctuations of a point on the original critical curve (i.e., $\tilde{\bm{\theta}}=(\delta\theta_{x},\theta_{y}+\delta\theta_{y})$) caused by substructures, the Jacobian matrix up to linear order is given by
\eq{
\lr{\frac{\partial \bm{\beta}(\tilde{\bm{\theta}})}{\partial \bm{\theta}}}\approx
\begin{pmatrix}
\epsilon \delta\theta_{x} - \delta\kappa -\delta \gamma_{1} & -\delta \gamma_{2} \\
-\delta \gamma_{2} & 2(1-\kappa_0)-\delta\kappa +\delta \gamma_{1} \\
\end{pmatrix},
}
where $\delta\kappa$, $\delta\gamma_{1}$, and $\delta\gamma_{2}$ represent the convergence and two components of the shear due to substructures, respectively. 
%show the components of the additional shear, which are defined by
%\eq{
%\delta\gamma_{1} = \frac{1}{2}(\psi_{,xx}-\psi_{,yy}), \quad \delta\gamma_{2} = \psi_{,xy}.
%}
The determinant of the Jacobian matrix is given by
\eq{
\mr{det}&\lr{\frac{\partial \bm{\beta}(\tilde{\bm{\theta}})}{\partial \bm{\theta}}}  \\
&\ \approx (\epsilon \delta\theta_{x} - \delta\kappa -\delta \gamma_{1})\lr{2(1-\kappa_0)-\delta\kappa +\delta \gamma_{1}}-\delta\gamma_{2}^2\\
&\ \approx 2(1-\kappa_0)(\epsilon\delta \theta_{x} - \delta\kappa -\delta \gamma_{1}).
}
Since the determinant must be equal to zero, the fluctuated critical curve satisfies
\eq{\label{eq: dtheta1_dkappa_dgamma1}
\delta \theta_{x} = \frac{1}{\epsilon}(\delta\kappa +\delta \gamma_{1}).
%, \quad \delta\theta_{y}=0.
}
% \KA{This equation could be derived for more general coordinate systems by noting that the following relationship holds near the critical curve,?}

From Eq.~\eqref{eq: dtheta1_dkappa_dgamma1}, we obtain the power spectrum of $\delta\theta_{x}$ as
\eq{\label{eq: Ptheta1_Pkappa_Pgamma1}
P_{\delta\theta_{x}}=\frac{1}{\epsilon^2}\lr{P_{\delta\kappa}+2P_{\delta\kappa\delta\gamma_{1}}+P_{\delta\gamma_{1}}},
}
where $P_{\delta\theta_{x}}$, $P_{\delta\kappa}$, and $P_{\delta\gamma_{1}}$ are the auto two-dimensional power spectrum of $\delta\theta_{x}$, $\delta\kappa$, and $\delta\gamma_{1}$, respectively, and $P_{\delta\kappa\delta\gamma_{1}}$ represents the cross power spectrum between $\delta\kappa$ and $\delta\gamma_{1}$.
Here, we define the two-dimensional power spectrum of $X$ and $Y$ as
\eq{
\braket{X(\bm{k}) Y(\bm{k}^\prime)}=(2\pi)^2\delta^{\mr{2D}}(\bm{k}+\bm{k}^\prime)P_{XY},
}
where $\bm{k}$ is a wave number on a two-dimensional plane.

Using the relations between $\delta\kappa$ and $\delta\gamma_1$,
\eq{
P_{\delta\gamma_{1}}=\cos^2(2\phi_\ell)P_{\delta\kappa},\\
P_{\delta\kappa\delta\gamma_{1}}=\cos(2\phi_\ell)P_{\delta\kappa},
}
we obtain
\eq{
P_{\delta\theta_{x}}=\frac{1}{\epsilon^2}(1+2\cos(2\phi_\ell)+\cos^2(2\phi_\ell))P_{\delta\kappa},
}
where $\phi_\ell$ is the azimuthal polar.
Taking average of $\phi_\ell$, we finally obtain
\eq{\label{eq: Pdtheta_Pdkappa}
P_{\delta\theta_{x}}=\frac{3}{2\epsilon^2}P_{\delta\kappa},
}
and
\eq{\label{eq: dtheta2_dkappa2}
%\frac{\braket{\delta\theta_x^2}}{\theta_{\mr{Ein}}^2} 
\epsilon^2\braket{\delta\theta_x^2}
= \frac{3}{2}\int d\log k~ \frac{k^2}{2\pi}P_{\delta\kappa} = \frac{3}{2}\braket{\delta\kappa^2}.
}
Note that $\epsilon^2\braket{\delta\theta_x^2} \approx \braket{\delta\theta_x^2}/\theta_{\mr{Ein}}^2$, given that $\epsilon \approx 1/\theta_{\mr{Ein}}$.
% \eq{\label{eq: dtheta2_dkappa2}
% \braket{\delta\theta_x^2} = \frac{3}{2\epsilon^2}\int d\log k~ \frac{k^2}{2\pi}P_{\delta\kappa} = \frac{3}{2\epsilon^2}\braket{\delta\kappa^2}.
% }
Deriving these simple formulas of Eqs~\eqref{eq: Pdtheta_Pdkappa} and~\eqref{eq: dtheta2_dkappa2} is the main result in this paper. 
Although previous work has derived a formula between fluctuations of image positions and the surface density perturbations from substructures~\cite{2002ApJ...572...25D}, as far as we are aware, this is the first time to derive the formulas between the critical curve fluctuations and the surface density perturbations.
These formulas allow us to analytically estimate the variance of $P_{\delta\theta_{x}}$ or $\braket{\delta\theta_x^2}^{1/2}$ from the surface density power spectrum of substructures, $P_{\delta\kappa}$.

While Eq.~\eqref{eq: Pdtheta_Pdkappa} is derived assuming a complete orthogonal coordinate system, we argue that Eq.~\eqref{eq: Pdtheta_Pdkappa} holds rather generically because near the fold critical curve tangential and radial magnifications generally behaves as~\cite{1986ApJ...310..568B}
\eq{
\mu_{\mathrm{t}}\approx \frac{\mu_{\mathrm{t0}}}{\delta\theta}, \quad \mu_{\mathrm{r}} \approx \mathrm{const}.,
}
where $\delta\theta$ denotes the distance from the critical curve. Since the tangential magnification is defined by $\mu_{\mathrm{t}}^{-1}=1-\kappa-\gamma$ with $\gamma=\sqrt{\gamma_1^2+\gamma_2^2}$, substructures modify the inverse of the tangential magnification as
\eq{
\mu_{\mathrm{t}}^{-1}\approx \frac{\delta\theta}{\mu_{\mathrm{t0}}}-\delta\kappa-\frac{\gamma_1}{\gamma}\delta\gamma_1-\frac{\gamma_2}{\gamma}\delta\gamma_2.
}
Since the perturbed critical curve satisfies $\mu_{\mathrm{t}}^{-1}=0$, we obtain
\eq{\label{eq: ptheta_pkappa_general}
P_{\delta\theta}&=\mu_{\mathrm{t0}}^2\left[P_{\delta\kappa}+\left(\frac{\gamma_1}{\gamma}\right)^2P_{\delta\gamma_1}+\left(\frac{\gamma_2}{\gamma}\right)^2P_{\delta\gamma_2}\right] \\
 &=\frac{3\mu_{\mathrm{t0}}^2}{2}P_{\delta\kappa},
}
where we take an average of $\phi_{\ell}$. Equation~\eqref{eq: ptheta_pkappa_general} is essentially same as Eq.~\eqref{eq: Pdtheta_Pdkappa} if $\mu_{\mathrm{t0}}=1/\epsilon$.
% which is essentially same as Eq.~\eqref{eq: Pdtheta_Pdkappa} if $\mu_{\mathrm{t0}}=1/\epsilon$. 

While our analytic results are applicable to any form of substructures,  here let us describe a specific form of $P_{\delta\kappa}(k)$ when substructures are CDM subhalos.
Using the halo formalism~\cite{2002PhR...372....1C} and assuming that the spatial correlation between subhalos can be negligible (i.e., subhalos are randomly distributed), we can compute the surface density power spectrum as an integral over the mass function weighted by their surface density profile as~\cite{2016JCAP...11..048H}
\eq{\label{eq: pkappa_sub}
P_{\delta\kappa}(k)=\int^{M_{\mr{max}}}_{M_{\mr{min}}} \frac{d n^{\mr{2D}}}{d M}\left|\tilde{\kappa}_M(k)\right|^2 d M,
}
where $k=\sqrt{k_x^2+k_y^2}$, $M_{\mr{min}}$ and $M_{\mr{max}}$ are the minimum and maximum mass of subhalos, respectively, $d n^{\mr{2D}}/d M$ is the surface number density of subhalos with masses of $[M, M+\delta M]$, and $\tilde{\kappa}_M(k)$ is the Fourier transform of the convergence $\kappa_M$ provided by a subhalo of mass $M$. Here $\tilde{\kappa}_M(k)$ can be calculated by
\eq{
\tilde{\kappa}_M(k)=\frac{M \tilde{u}_M(\bm{K}_*=(k_x,k_y,0))}{\Sigma_{\rm cr}},
}
where $\Sigma_{\rm cr}$ is the critical surface density, and $\tilde{u}_M(\bm{K})$ is the Fourier transform of their three-dimensional density profile, $\rho_M$, (see the Appendix), 
\eq{
\tilde{u}_M(\bm{K}) &= \int_0^{R_{\mr{vir}}} \frac{4 \pi R^2}{M} \frac{\sin KR}{KR} \rho_M(R) \mathrm{d} R,
}
where $K=|\bm{K}|$.
% \eq{
% \tilde{u}_M(K) &= \int_0^{r_{v i r}} \frac{4 \pi r^2}{M} \frac{\sin k r}{k r} \rho_M(r) \mathrm{d} r.
% }
In this paper, we simply adopt the ordinal Navarro-Frenk-White~(NFW) profile~\cite{1996ApJ...462..563N},
which is defined by 
\eq{\label{eq: nfw_profile}
\rho_{\mr{NFW}}(R) = \frac{M c_{\mr{vir}}^3}{4\pi R_{\mr{vir}}^3}\frac{f_{\mr{NFW}}(c_{\mr{vir}})}{\left(R c_{\mr{vir}}/ R_{\mr{vir}}\right)\left(1+R c_{\mr{vir}}/ R_{\mr{vir}}\right)^2},
}
% \eq{\label{eq: nfw_profile}
% \rho_{\mr{NFW}}(r) = \frac{M c_{\mr{vir}}^3}{4\pi r_{\mr{vir}}^3}\frac{f_{\mr{NFW}}(c_{\mr{vir}})}{\left(r c_{\mr{vir}}/ r_{\mr{vir}}\right)\left(1+r c_{\mr{vir}}/ r_{\mr{vir}}\right)^2},
% }
where $R_{\mr{vir}}$ is the virial radius, $c_{\mr{vir}}$ is the concentration parameter for virial radius, and $f_{\mr{NFW}}(c) \equiv 1/[\ln (1+c)-c/(1+c)]$.
Then $\tilde{u}_M(K)$ can be expressed by~\cite{2001ApJ...546...20S},
\eq{
\tilde{u}_M(K) &=f_{\mr{NFW}}(c_{\mr{vir}})\Biggl\{\sin \mathcal{K}[\operatorname{Si}(\mathcal{K}(1+c_{\mr{vir}}))-\operatorname{Si}(\mathcal{K})]  \\
& ~+\cos \mathcal{K}[\operatorname{Ci}(\mathcal{K}(1+c_{\mr{vir}}))-\operatorname{Ci}(\mathcal{K})]-\frac{\sin (\mathcal{K} c_{\mr{vir}})}{\mathcal{K}(1+c_{\mr{vir}})}\Biggr\},
}
with $\mathcal{K}\equiv KR_{\mr{vir}}$, the sine integral functions, $\operatorname{Si}(X)=\int_0^X d t \sin (t) / t$, and the cosine integral function, $\operatorname{Ci}(X)=-\int_X^{\infty} d t \cos (t) / t$.

\section{Validity of the formula connecting $\braket{\delta\theta^2}$ and $\braket{\delta\kappa^2}$}\label{sec: validity_cc}
In the previous section, we derive Eqs.~\eqref{eq: Pdtheta_Pdkappa} and~\eqref{eq: dtheta2_dkappa2}, the simple analytical relations between fluctuations of the macrocritical curve and the fluctuation of the surface density due to substructures.
Here, we check the validity of Eq.~\eqref{eq: dtheta2_dkappa2} by comparing it with direct numerical estimations of fluctuations of macrocritical curves due to substructures.

We first set up macrolens models.
We assume that the macrolens object is a DM halo with a mass of $M_{\mr{hh}}=10^{15}~M_{\odot}/h$ located at redshift  $z_\mr{l}=0.5$.
We here focus on the tiny surface area near the critical curve caused by the DM halo adopting the lens potential defined by Eq.~\eqref{eq: deflection_potential} with $\epsilon = 1/\as{10}$ and the convergence $\kappa_0=0.5$. 

We next set up lens models of substructures hosted by the DM halo. As an example, we here consider CDM subhalos as substructures. 
For the mass distribution of subhalos, we adopt the NFW profile in Eq.~\eqref{eq: nfw_profile} with the concentration parameter computed by the mass-concentration relation presented in Ref.~\cite{2020MNRAS.492.3662I}. Note that we multiply the concentration parameter by $\left[200 / \Delta_{\mathrm{vir}} \Omega_{\mathrm{m}}(z)\right]^{1 / 3}\left[H(z) / H_0\right]^{-2 / 3}$ with the virial overdensity computed from the spherical collapse model, $\Delta_{\mathrm{vir}}$, to convert from $c_{200}$ to $c_{\mathrm{vir}}$~\cite{10.1143/PTP.97.49}.

We also assume that the density profile of the DM halo follows the NFW density profile, and the spatial distribution of subhalos also follows it. The surface number density of subhalos is then proportional to the surface NFW density as
\eq{
n^{\mr{2D}}\equiv \frac{dN_{\mr{sub}}}{dS} \propto \Sigma_{\mr{NFW}}(r;M_{\mr{hh}}),
}
where $dS = 2\pi rdr$ with $r=\sqrt{x^2+y^2}$, $\Sigma_{\mr{NFW}}(r)$ is the surface density of the NFW profile. The concentration parameter for the host halos is estimated by the mass-concentration relation presented in Ref.~\cite{2015ApJ...799..108D}.
The surface number density of subhalos with masses of $[M, M+\delta M]$ in Eq.~\eqref{eq: pkappa_sub} is given by
\eq{\label{eq: dn2ddM_sub}
\frac{dn^{\mr{2D}}}{dM}=\frac{dN_{\mr{sub}}}{dM}\frac{\Sigma_{\mr{NFW}}(r;M_{\mr{hh}})}{M_{\mr{hh}}},
}
where $dN_{\mr{sub}}/dM$ shows the total number of subhalos with masses of $[M, M+\delta M]$ inside their host halo.
% Since we here focus on the tiny surface Considering the surface number density with masses of $[M, M+\delta M]$, one can obtain
% \eq{
% \frac{d n^{\mr{2D}}}{d M} \approx 2\frac{d_{\mr{A}}}{\epsilon}\frac{dN_{\mr{sub}}}{dM}\frac{\rho_{\mr{NFW}}(r=d_{\mr{A}}/\epsilon;M_{\mr{hh}})}{M_{\mr{hh}}},
% }
% where $\frac{dN_{\mr{sub}}}{dM}$ shows the total number of subhalos with masses of $[M, M+\delta M]$ inside their host halo. 
We adopt a simple analytic model presented in Ref.~\cite{2020ApJ...901...58O} for calculating $dN_{\mr{sub}}/dM$.
We assume that the spatial correlation among subhalos can be neglected and distribute them with a constant surface number density assuming the Poisson distribution, which is justified by the tiny surface area in the vicinity of the critical curve considered here.
This assumption may be validated by the fact that subhalos are exposed to the tidal gravity of their host halo.

Distributing CDM subhalos near the macrocritical curve by the Poisson distribution with the expected number from Eq.~\eqref{eq: dn2ddM_sub}, we numerically calculate fluctuations of the macrocritical curves.
We use an open software, {\tt Glafic} code, presented in Refs.~\cite{2010PASJ...62.1017O,2021PASP..133g4504O}.
To distribute subhalos, it is necessary to set $M_{\mr{min}}$ and $M_{\mr{max}}$ and the box size considered here. 
We explore 12 models with different sets of $(M_{\mr{min}}, M_{\mr{max}})$ summarized in Table~\ref{tab: results}.
The box size is basically set to $\as{6.0}\times \as{6.0}$ with the exception of $\as{3.0}\times \as{3.0}$ for model (iii).
These box sizes are determined for numerical reasons.

\begin{table*}[ht]
 \caption{Our models with different $(M_{\mr{min}}, M_{\mr{max}})$ for the validation. We show numerical results as well as analytically estimated values of $\braket{\theta_{x}^2}^{1/2}$ by Eq.~\eqref{eq: dtheta2_dkappa2} for individual models.}
 \label{tab: results}
  \centering
  \begin{tabular}{cccccc}
   \hline
   Model & $M_{\mr{min}}~[\Ms/h]$ & $M_{\mr{max}}~[\Ms/h]$ & Numerical results~[\as{}] & Analytic value~[\as{}] & Remark\\
   \hline \hline
   (i) & $5\times 10^{6}$ & $5\times 10^7$ & $0.0334 \pm 0.0055$ & $0.0342$ & $\as{6.0}\times \as{6.0}$, 10 realizations \\
   (ii) & $5\times 10^{6}$ & $10^8$ & $0.0459 \pm 0.0123$ & $0.0422$ & $\as{6.0}\times \as{6.0}$, 10 realizations \\
   (iii) & $10^{6}$ & $10^8$ & $0.0476 \pm 0.0114$ & $0.0467$ & $\as{3.0}\times \as{3.0}$, 10 realizations \\
   (iv) & $5\times 10^{6}$ & $5\times 10^8$ & $0.0631 \pm 0.0108$ & $0.0630$ & $\as{6.0}\times \as{6.0}$, 10 realizations \\
   (v) & $5\times 10^{6}$ & $10^9$ & $0.0629 \pm 0.0141$ & $0.0730$ & $\as{6.0}\times \as{6.0}$, 10 realizations \\
   (vi) & $5\times 10^{6}$ & $5\times 10^9$ & $0.113 \pm 0.0391$ & $0.103$ & $\as{6.0}\times \as{6.0}$, 10 realizations \\
   (vii) & $5\times 10^{6}$ & $10^{10}$ & $0.0995 \pm 0.0320$ & $0.119$ & $\as{6.0}\times \as{6.0}$, 10 realizations \\ 
   (viii) & $5\times 10^{6}$ & $5\times 10^{10}$ & $0.150 \pm 0.055$ & $0.163$ & $\as{6.0}\times \as{6.0}$, 10 realizations \\
   (ix) & $5\times 10^{6}$ & $6\times 10^{10}$ & $0.122 \pm 0.044$ & $0.170$ & $\as{6.0}\times \as{6.0}$, 10 realizations \\ 
   % (i) & $5\times 10^{6}$ & $8\times 10^{10}$ & $0.150 \pm 0.0553$ & $0.1627$ & $\as{3.0}\times \as{3.0}$, 10 realizations \\ 
   (x) & $5\times 10^{6}$ & $10^{11}$ & $ 0.112 \pm  0.044$ & $0.186$ & $\as{6.0}\times \as{6.0}$, 10 realizations \\ 
   (xi) & $5\times 10^{6}$ & $5\times 10^{11}$ & $ 0.122 \pm 0.0529$ & $0.256$ & $\as{6.0}\times \as{6.0}$, 10 realizations \\ 
   (xii) & $5\times 10^{6}$ & $10^{12}$ & $0.129\pm 0.041$ & $0.291$ & $\as{6.0}\times \as{6.0}$, 10 realizations \\ \hline
  \end{tabular}
\end{table*}

\begin{figure*}[t]
    \begin{tabular}{ccc}
       \begin{minipage}{.32\linewidth}
        \centering
        \includegraphics[width=.9\hsize]{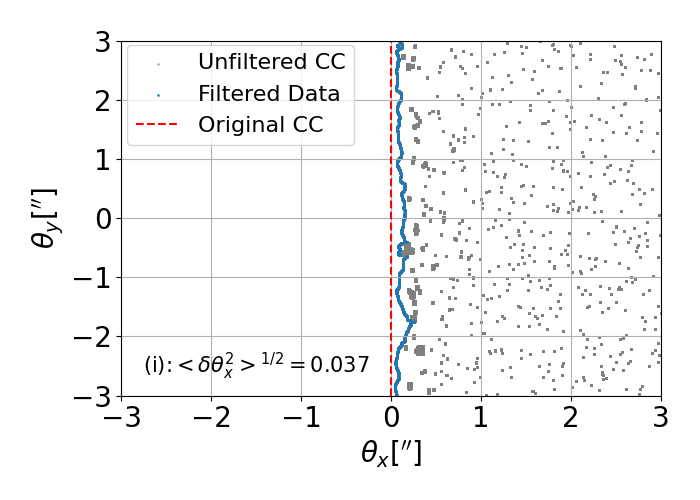}
      \end{minipage} &
      \begin{minipage}{.32\linewidth}
        \centering
        \includegraphics[width=.9\hsize]{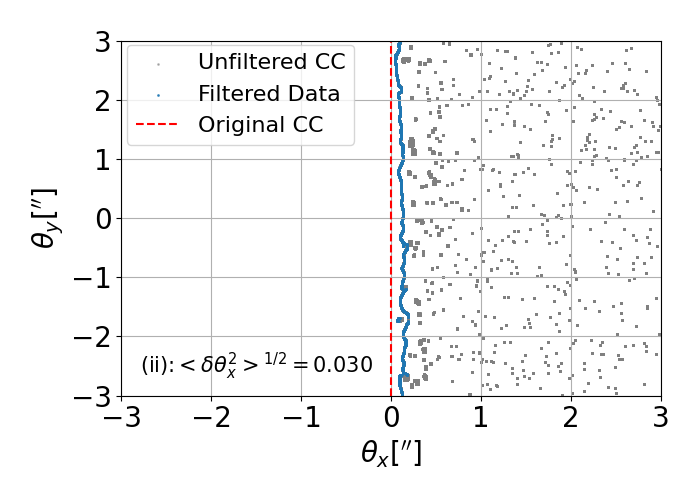}
      \end{minipage} &
   
      \begin{minipage}{.32\linewidth}
        \centering
        \includegraphics[width=.9\hsize]{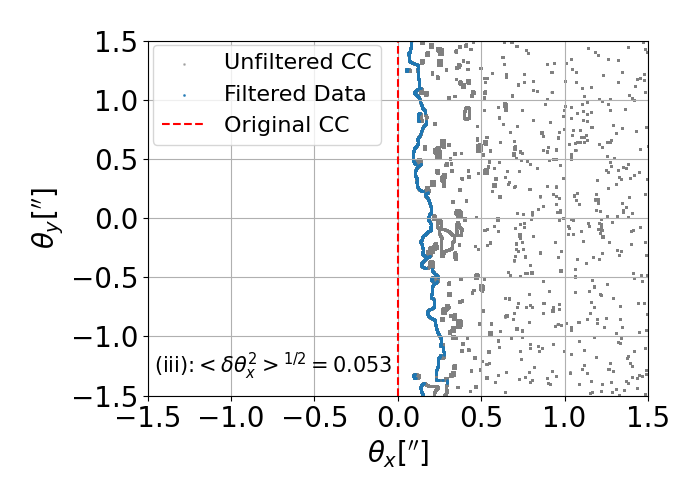}
      \end{minipage}\\
      \begin{minipage}{.32\linewidth}
        \centering
        \includegraphics[width=.9\hsize]{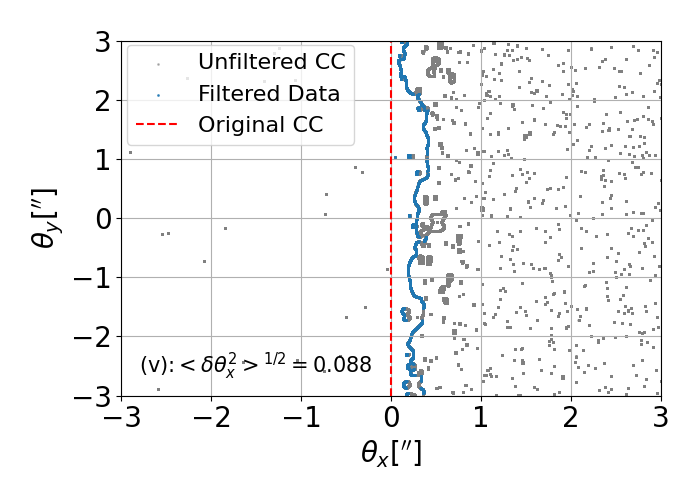}
    \end{minipage} &
   
      \begin{minipage}{.32\linewidth}
        \centering
        \includegraphics[width=.9\hsize]{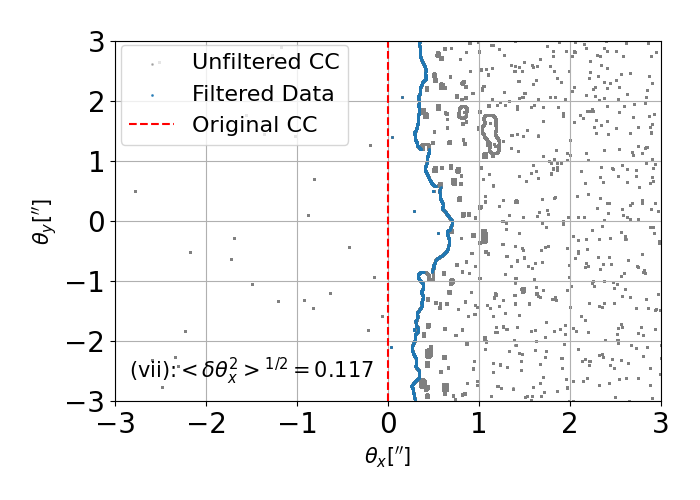}
      \end{minipage} &
      \begin{minipage}{.32\linewidth}
        \centering
        \includegraphics[width=.9\hsize]{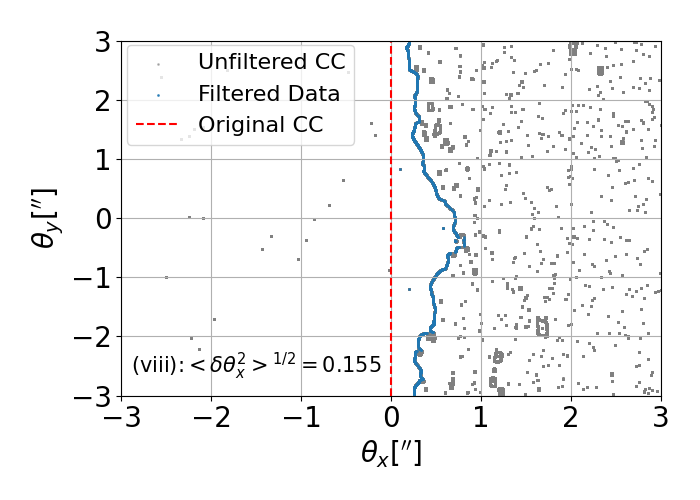}
    \end{minipage}\\
      \begin{minipage}{.32\linewidth}
        \centering
        \includegraphics[width=.9\hsize]{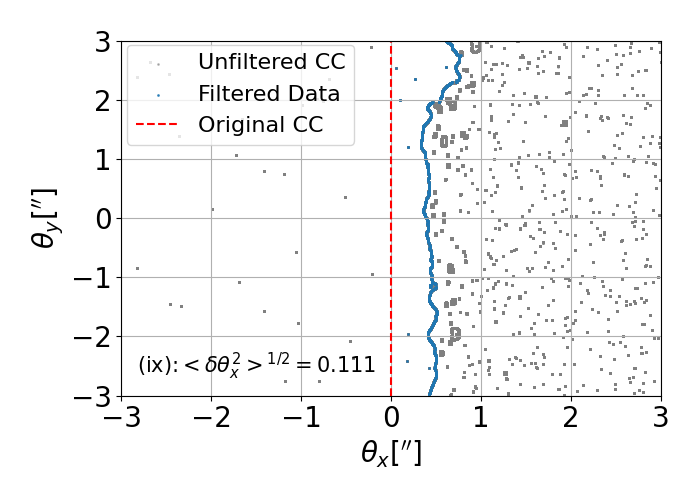}
    \end{minipage} &
   
      \begin{minipage}{.32\linewidth}
        \centering
        \includegraphics[width=.9\hsize]{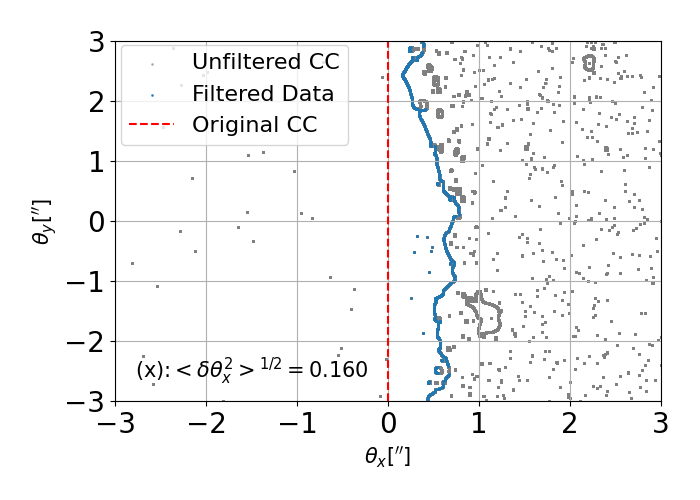}
      \end{minipage} &
      \begin{minipage}{.32\linewidth}
        \centering
        \includegraphics[width=.9\hsize]{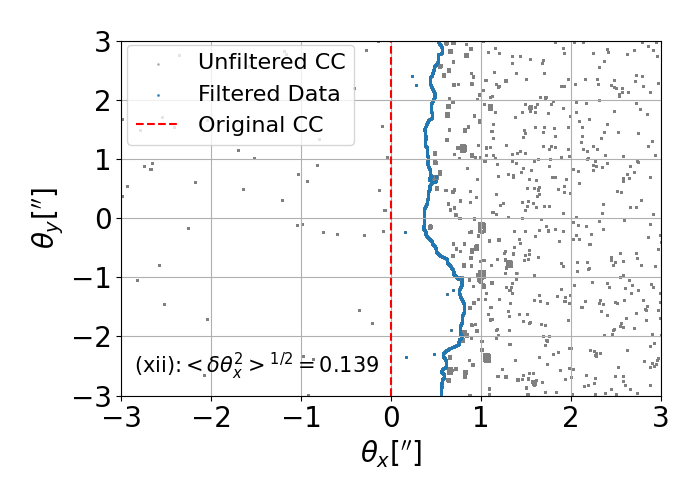}
    \end{minipage}
    \end{tabular}
     \caption{An example of fluctuated critical curves by CDM subhalos for each model summarized in Table~\ref{tab: results}. The red dashed vertical line shows the original macrocritical curve described in Sec.~\ref{sec: fluct_cc}. The gray points show critical curves calculated in {\tt Glafic}, while the blue points show the fluctuated critical curves obtained by filtering, as explained in the text. We also show the value of $\braket{\delta\theta_x^2}^{1/2}$ for each panel in units of arcsec. For the mean value of each model, see Table~\ref{tab: results}.
     Top left: model (i), Top center: model (ii), Top right: model (iii), Middle left: model (v), Middle center: model (vii), Middle right: model (viii), Bottom left: model (ix), Bottom center: model (x), Bottom right: model (xii).}
     \label{fig: cc_each_model}
     
  \end{figure*}

\begin{figure}[t]
    \includegraphics[width=8cm,clip]{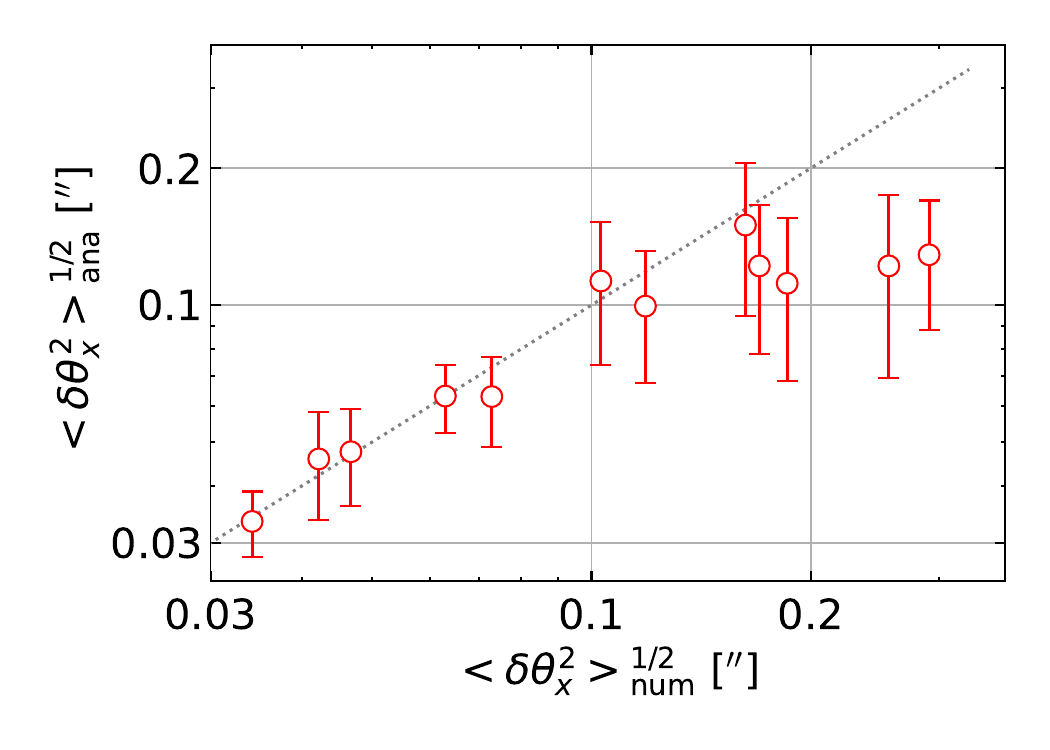}
    \centering
    \caption{Comparison between analytically calculated values of $\braket{\delta\theta_x^2}^{1/2}$ from Eq.~\eqref{eq: dtheta2_dkappa2} and the ones numerically estimated using {\tt Glafic}. Here we set $M_\mr{hh}=10^{15}~\Ms/h$.}
    \label{fig: cor_dthetax}
\end{figure}

For each model of $(M_{\mr{min}}, M_{\mr{max}})$, we run the calculations for ten realizations.
From the obtained critical curves, we take the mean and variance of the fluctuations $\braket{\theta_{x}^2}^{1/2}$.
A caveat is that {\tt Glafic} outputs isolated critical curves that are associated with subhalos as well.
In order to compute fluctuations of macromodel critical curves that are our main interest here, those critical curves associated with subhalos must be removed. To do so, we adopt a simple filtering method with three steps:~(1)~sorting points representing critical curves by $\theta_y$ coordinate, (2)~subdividing the sorted points into several blocks with $\mathcal{O}(10^2)$ data points, and (3)~acquiring only the point with the smallest $\theta_x$ in each block.
Figure~\ref{fig: cc_each_model} shows the resulting fluctuated critical curves for each model.
In Fig.~\ref{fig: cc_each_model}, the red dashed line shows the original macrocritical curve described in Sec.~\ref{sec: fluct_cc}. 
The gray points depict critical curves calculated in {\tt Glafic}.
The blue points show the fluctuated critical curves obtained by the filtering method described above.
We find that the original critical curves are shifted in the positive direction of the $\theta_x$axis. This is explained by the contribution of the additional mass of subhalos that moves the macro critical curve to the outer radii, which corresponds to the position $\theta_x$ in this coordinate system.

Table~\ref{tab: results} also shows the numerical results of $\braket{\theta_{x}^2}^{1/2}$ estimated in the manner described above. We find that the values of $\braket{\theta_{x}^2}^{1/2}$ for models (i) - (viii) agree well with the analytically estimated values from Eq.~\eqref{eq: dtheta2_dkappa2}. 
% When the $M_{\mr{max}}$ becomes larger than $10^{10}M_{\odot}$, the value deviates from the analytic values.
When the $M_{\mr{max}}$ becomes larger than $5\times10^{10}M_{\odot}$, the numerically estimated values become smaller than the analytically estimated ones.
This deviation may come from the non-Gaussianity due to massive subhalos; the probability of their existence is rare, i.e., there seldom exists such massive subhalos in our calculation box, while their analytical contribution to $\braket{\kappa^2}$ is significant. Thus, examining whether the distribution of $\delta \theta_x$ is Gaussian or not should provide useful guidance on this deviation. In addition, we may think of this saturation as occurring in models that contain subhalos so massive that none of them are present in our computational region in many realizations. Since we perform our calculations for the region of $10\times (\as{6}\times \as{6})$, this mass scale corresponds to $\sim 10^{10} M_{\odot}$, which indicates the deviation at the scale of $\braket{\delta \theta_x^2}^{1/2}\gtrsim \as{0.1}$.
Moreover, our filtering method to pick the smallest-$\theta_x$ data point in each block could lead to systematic underestimations in some simulations. In any case, we conclude that Eq.~\eqref{eq: dtheta2_dkappa2} is valid and accurate, at least as long as the substructure power spectrum is not dominated by a small number of massive structures. We summarize the comparison between the numerical and analytic values from Eq.~\eqref{eq: dtheta2_dkappa2} in Fig.~\ref{fig: cor_dthetax}.

\section{Application to Mothra}\label{sec: mothra}
This section will apply the formula in Eq.~\eqref{eq: dtheta2_dkappa2} for discussing the origin of an extremely magnified binary star at redshift $z=2.091$ recently reported in Ref.~\cite{2023arXiv230710363D} with JWST/NIRCam data, nicknamed ``Mothra.''
Mothra is found in the strong lensing region in the galaxy cluster of \textrm{MACS J0416.1-403} at $z =0.397$.
The Mothra, formally called LS1, is observed only on one side of the critical curve with negative parity, and its counterimage is not seen even though nearby star clusters are found in a pair on both sides of the critical curve as shown in Fig.~2 of Ref.~\cite{2023arXiv230710363D}.
% The ``Mothra" is focused as an anomaly binary star, formally called LS1~(hereafter we refer to as LS1), because the binary star is observed only on the side of the critical curve with negative parity, and the counterimage is not seen yet even though nearby stars are found in set with their counterimages as shown in Fig.~\ref{fig: mothra}.
% LS1 is informed to be in a galaxy with spectroscopic redshift $z=2.091$ and in a portion of the galaxy that is parsecs away from the cluster caustic. 
% The macrolensing is caused by the galaxy cluster of MACS J0416.1−403 (“M0416”) at $z =0.397$, which has been observed for the abundant detection of lensed stars.
%Mothra is informed to be in a galaxy with spectroscopic redshift $z=2.091$. 

Although Ref.~\cite{2023arXiv230710363D} argued that the anomaly could be explained by considering a local millilensing effect due to a substructure that demagnifies and hides one of the multiple images of Mothra, we here attempt to interpret this anomaly as caused by fluctuations in the macrocritical curve due to substructures. In our new interpretation, Mothra is regarded as an event similar to Earendel~\cite{2022Natur.603..815W} that is located exactly on the macrocritical curve and whose multiple images are unresolved, but due to a fluctuated macrocritical curve Mothra is observed in an apparently offset position. With our analytic formula, we check which substructure models can explain the observed offset of Mothra.
% , unlike the main scenario proposed in Ref.~\cite{2023arXiv230710363D}.

% \begin{figure}[htbp]
%     \centering
%     \includegraphics[width=8cm,height=6cm,clip]{figs/mothra.png}
%     \caption{Enlarged image of Mothra (shown as LS1) and its surroundings. Three multiply lensed knots near Mothra and their counterimages are also labelled. However, there is no visible counterimage for Mothra. While the solid white curve is the expected position of the critical curve based on the macrolens model, the white dashed line shows the inferred position based on the pairs of (b) and (c) images. We assume that the later position is correct and consider the fluctuation from the white dashed line. This figure is taken from Ref.~\cite{2023arXiv230710363D}.}
%     \label{fig: mothra}
% \end{figure}

The macrolens model for the Mothra is set up as follows. We adopt $z_\mr{s}=2.091$ and $z_{\mr{l}}=0.397$.
According to Ref.~\cite{2014ApJ...795..163U}, the virial halo mass of \textrm{MACS J0416.1-403} is estimated as $(1.24 \pm 0.28)\times 10^{15}\Ms$.
%which is defined by a virial overdensity based on the spherical collapse model~\cite{10.1143/PTP.97.49}.
Using the {\tt Glafic} mass model of \textrm{MACS J0416.1-403}~\cite{2010PASJ...62.1017O,2016ApJ...819..114K}, the Einstein radius for this source redshift is estimated as $\theta_{\mr{Ein}}\approx \as{24.13}$.
%The size of the galaxy cluster at the major axis is about $\as{104.72}$, while the one at the minor axis is about $\as{22.24}$.
%Then, we fix the Einstein radius to $\theta_{\mr{Ein}}\approx \sqrt{\as{52.36}\times \as{11.12}} = \as{24.13}$.
Combining the Einstein radius and the mass of the galaxy cluster, we also fix the concentration parameter $c_{\mr{vir}}=7.59$. Here, we assume the stellar-mass-halos-mass relation~\cite{2019MNRAS.488.3143B}, galaxy-size relation~\cite{2020ApJ...901...58O}, the Hernquist density profile for the stellar components~\cite{1990ApJ...356..359H}, and the mass-concentration relation~\cite{2015ApJ...799..108D}. 
Note that the total stellar mass is estimated as $1.09\times 10^{12}\Ms$, and the effective radius in the Hernquist profile is $\theta_\mr{b}=\as{1.22}$.
The parameter $\epsilon$ should be determined by the local structure of the macromodel critical curve around Mothra. Again, we adopt the {\tt Glafic} mass model to estimate the tangential magnification as a function of the distance from the macrocritical curve $\delta\theta$, finding $\mu_{\mathrm{t}}\approx \as{8}/\delta\theta$. Based on the discussion given in Sec.~\ref{sec: fluct_cc}, we set $\epsilon=1/\as{8}$ for our analytic estimates of fluctuations of macrocritical curves.

In our new interpretation, the macrocritical curve needs to be fluctuated by $\sim \as{0.07}$, especially to the negative parity side, as shown in Fig.~2 in Ref.~\cite{2023arXiv230710363D}.
% To trigger the anomaly of Mothra, the macro critical curve needs to be fluctuated by $\sim 0^\as.07$, especially to the negative parity side, as shown in Fig.~\ref{fig: mothra}.
Equation~\eqref{eq: dtheta2_dkappa2} allows us to analytically estimate the fluctuation of the surface density power spectrum by substructures necessary to predict the fluctuation of the macrocritical curve needed to explain Mothra.
We obtain $\braket{\delta\kappa^2}\sim 3.27\times 10^{-5}$ for $\braket{\delta\theta_x^2}^{1/2}\sim \as{0.07}$.
In the following, we consider two cases as substructures as examples: CDM subhalos and quantum clumps of FDM halos.

\subsection{CDM subhalos}
\begin{figure}[t]
    \centering
    \includegraphics[width=8cm,clip]{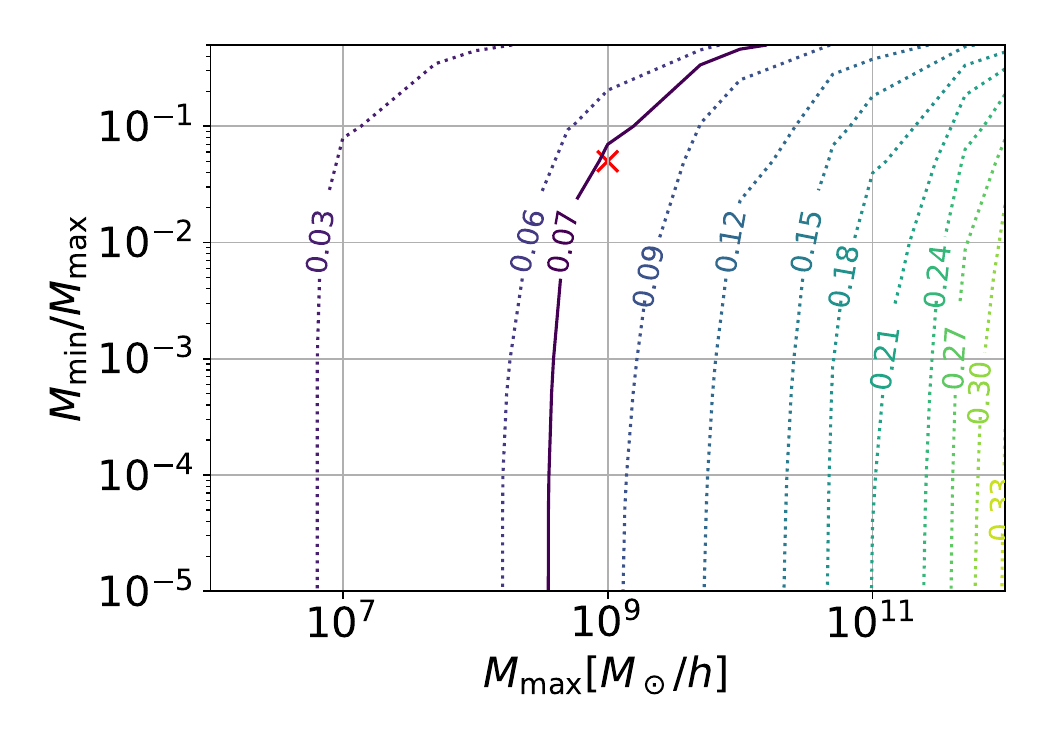}
    \caption{Contour of $\braket{\delta\theta_x^2}^{1/2}$ as a function of $M_{\mr{max}}$ and $M_{\mr{min}}/M_{\mr{max}}$ for the Mothra-like lens system. The value of each contour is shown in units of arcsec. The red cross shows the model that we calculate in detail to interpret the Mothra's anomalous position.}
    \label{fig: dthetax_contour_mothra}
\end{figure}

Combining Eq.~\eqref{eq: pkappa_sub} with Eq.~\eqref{eq: dtheta2_dkappa2}, we estimate the parameter set of ($M_{\mr{min}}$, $M_{\mr{max}}$) needed to explain $\braket{\delta\kappa^2}\sim 3.27\times 10^{-5}$.
% $\braket{\theta_x^2}^{1/2}\sim 0.07^\as$, so that $\braket{\delta\kappa^2}\sim 3.27\times 10^{-5}$.
% Using the relation in Eq.~\eqref{eq: dtheta2_dkappa2}, we can easily predict $\braket{\delta\theta_x^2}^{1/2}$ as a function of $M_{\mr{max}}$ and $M_{\mr{min}}/M_{\mr{max}}$ for the lens system. 
%Analytical calculation of Eqs.~\eqref{eq: dtheta2_dkappa2} and~\eqref{eq: pkappa_sub} gives 
We plot the contour of $\braket{\delta\theta_x^2}^{1/2}$ as a function of $M_{\mr{max}}$ and $M_{\mr{min}}/M_{\mr{max}}$ in Fig.~\ref{fig: dthetax_contour_mothra}. 
Since the minimum mass of CDM subhalos is actually very small, e.g., $10^{-12}-10^{-3}\Ms$ for the supersymmetric neutralino~\cite{2001PhRvD..64h3507H,2004MNRAS.353L..23G,2006PhRvL..97c1301P,2015PhRvD..92f5029D}, the $M_\mr{min}/M_\mr{max}$ ratio would be almost zero. From Fig.~\ref{fig: dthetax_contour_mothra}, the value of $M_\mathrm{max}$ to explain the Mothra-like lens system becomes smaller than $10^9~M_\odot/h$ for such very small $M_\mr{min}/M_\mr{max}$ ratio, which is consistent with the fact that there is no visible galaxy around Mothra. Thus, we find that CDM subhalos, especially with $M_{\mr{max}}\sim 10^8-10^9~M_\odot/h$, have the potential to explain the Mothra-like lens system.

In addition to the Mothra, there are several multiple image pairs of star clusters on both sides of the critical curves. CDM subhalos considered in this paper may also affect the positions and magnification ratios of these multiple image pairs of star clusters. 
We further check the validity of our model shown in Fig.~\ref{fig: dthetax_contour_mothra} by exploring whether CDM subhalos significantly affect the magnification ratios of those multiple image pairs, focusing on the image pair $c$ and $c'$ defined in Ref.~\cite{2023arXiv230710363D}\footnote{We have confirmed that Eq.~\eqref{eq: dtheta2_dkappa2} holds for the model as well.}.
We here choose a model with $(M_{\mr{min}}, M_{\mr{max}})=(5\times 10^7~\Ms/h, 10^9~\Ms/h)$ as an example that potentially explains the Mothra-like lens system for a numerical reason. This model is represented in Fig.~\ref{fig: dthetax_contour_mothra} by the red cross.
% further to investigate whether such fluctuations could affect the magnification ratio between $c$ and $c^\prime$, a nearby multiple image pair of a star cluster.
% As mentioned in Ref.~\cite{2023arXiv230710363D}, the magnification ratio should approximately correspond to the ratio of separations from $b$ to $c$ and from $b^\prime$ to $c^\prime$, $\sim (b-c)/(b^\prime-c^\prime) \sim 0.39/0.32 = 1.22$.
Reference~\cite{2023arXiv230710363D} suggests that the ratio of the magnification at the position of the image $c$ against the one of the image $c^\prime$ should approximately be $1.22$ from the configurations of these multiple image pairs.
We numerically calculate fluctuated critical curves for this model in the same manner as in Sec.~\ref{sec: validity_cc}. In the calculation, we here also record the magnification at positions of the image $c$, $\mu_{c}$, and of the image $c^\prime$, $\mu_{c^\prime}$. Running the calculations in ten realizations for each model, we take the 2000 parameter sets of $(\delta\theta_x, \mu_c, \mu_{c^\prime}).$
% We numerically calculate fluctuated critical curves for this model in the same manner as in Sec.~\ref{sec: validity_cc} by subdividing the data points and collecting only the point with the smallest $\theta_x$ in each block and, at the same time, record the values of the magnification $\mu$ at positions of $c$ and $c^\prime$ in each block. Running the calculations in ten realizations for each model, we take the 2000 parameter sets of $(\delta\theta_x, \mu_c, \mu_{c^\prime}).$

\begin{figure}[t]
    \captionsetup{justification=justified,singlelinecheck=false}
    \includegraphics[width=8cm,height=6cm,clip]{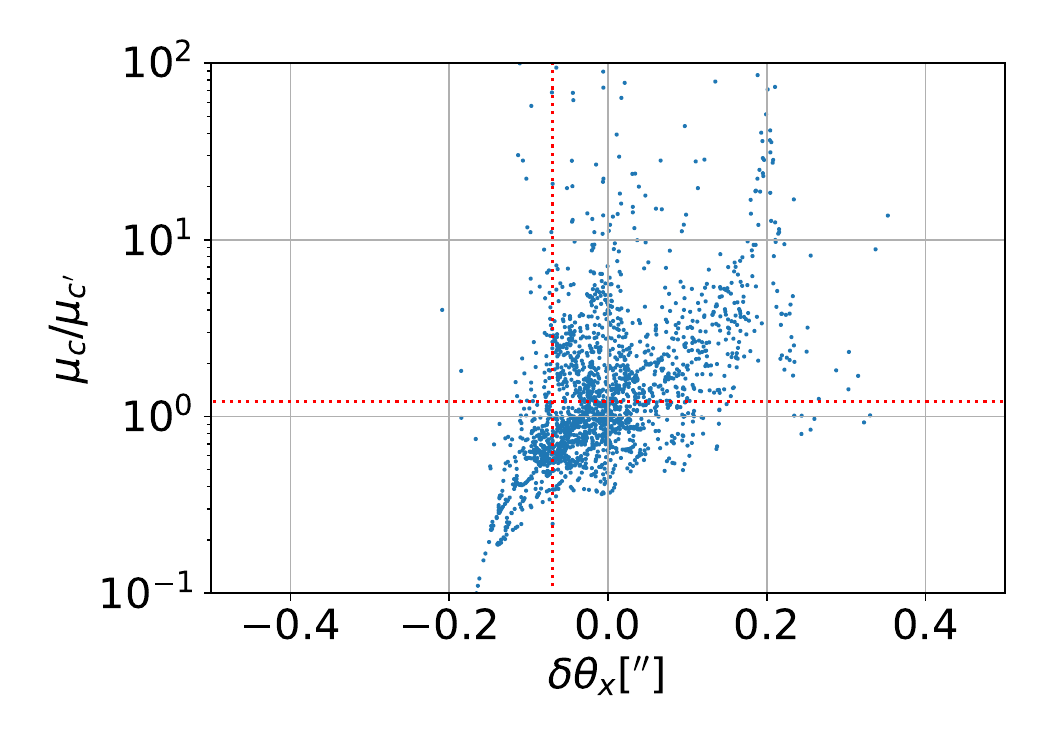}
    \caption{The correlation between $\delta\theta_x$ and the magnification ratio between $c$ and $c^\prime$, multiple image pair near Mothra. 
    %We set the halo mass $M_{\rm h}$ $=$ $1.24 \times 10^{15}\ M_{\odot}$, the stellar mass $M_{\rm s}$ $=$ $1.09 \times 10^{12}\ M_{\odot}$, the halo concentration $c_{h, \rm vir}$ $=$ $7.59$, the effective radius in Hernquist profile $\theta_{\rm b} = 1.22"$ and Einstein radius $\theta_{\rm Ein} = 24.13"$. 
    From the configuration of multiple images, the magnification ratio is predicted to be about $1.22$, represented by the horizontal red dotted line. The vertical red line shows $\delta\theta_x$ to make the image at the Mothra's offset an unresolved pair of images.}
    \label{fig: dthetax_mag_ratio_mothra}    
\end{figure}

Figure~\ref{fig: dthetax_mag_ratio_mothra} shows the correlation between $\delta\theta_x$ and the magnification ratio $\mu_c/\mu_{c^\prime}$. The blue points show the parameter sets obtained in our calculation, while the red dotted vertical line indicates the value of $\delta\theta_x$ required to reproduce the Mothra-like lens system.
% The blue points show each parameter set, while the red dotted vertical line indicates the approximate required value of $\delta\theta_x$ to make the critical curve pass through the Mothra position. 
The red dotted horizontal line shows the magnification ratio $\mu_c/\mu_{c^\prime}=1.22$ as suggested in Ref.~\cite{2023arXiv230710363D}.
% The red dotted horizontal line shows the magnification ratio at the position of $c$ and $c^\prime$ expected from the distance ratio between $b-c$ and $b^\prime-c^\prime$.
When $|\delta\theta_x|\sim \as{0.2}$, the fluctuated critical curve passes near the position of the image $c$ or the image $c^\prime$, which leads to the strong correlation between $\delta\theta_x$ and $\mu_c/\mu_{c^\prime}$ as shown in Fig.~\ref{fig: dthetax_mag_ratio_mothra}. However, we find that $\mu_c/\mu_{c^\prime}$ does not strongly correlate with $\delta\theta_x$ when $\delta\theta_x\lesssim \as{0.1}$, indicating the possibility of reproducing the Mothra-like lens-system while keeping the observed magnification ratio of $\mu_c/\mu_{c^\prime}\sim 1.22$.

It should be noted that we also need to consider perturbations on the image positions of several multiple image pairs of star clusters around Mothra due to subhalos to confirm that those subhalos do not significantly affect the observed positions of those additional multiple images. We leave them for future research.

\subsection{Quantum clumps in FDM halos}
Due to the wave nature of FDM, the quantum clumps (granular structures), which originate from the interference pattern, are observed in FDM halos.
The size of each clump corresponds to the de Broglie wavelength. 
The surface density perturbations due to these clumps are analytically studied in Ref.~\cite{2022ApJ...925...61K},
% Here the sub-galactic matter power spectrum is calculated under the assumptions that the clumps are randomly distributed with the number density $\langle n\rangle = 1/V_{\rm c}$, where $V_{\rm c}$ is the (constant) volume of each clump, and that the mass of each clump is determined by the local NFW density, and that the density profile of each clump is gaussian.
in which the subgalactic matter power spectrum is calculated under the assumptions that the clumps are randomly distributed with the number density $\langle n\rangle = 1/V_{\rm c}$, where $V_{\rm c}$ is the (constant) volume of each clump, and that the mass of each clump is determined by the local NFW density.
In addition, the density profile of each clump is assumed to be Gaussian.

% Without the smooth baryon profile, the surface density power spectrum can be calculated as
Without baryon components, the surface density power spectrum can be calculated as
\begin{equation}
    P_{\delta\kappa, {\rm FDMonly}}(k) = \frac{\pi \int dz\ \rho_{\rm NFW}^{2}(R)}{6\Sigma_{\rm cr}^{2}} \lambda_{\rm c}^{3}\exp \left(-\frac{\lambda_{\rm c}^{2}k^{2}}{4} \right), \label{eq:Pkappa_fdmonly}
\end{equation}
where $\lambda_{\rm c}$ is the de Broglie wavelength.
% where $\lambda_{\rm c}$ is the de Broglie wavelength and $\Sigma_{\rm cr}$ is the critical surface density.
From Eqs.~\eqref{eq: Pdtheta_Pdkappa} and \eqref{eq: dtheta2_dkappa2}, we can estimate the critical curve perturbation in FDM halos as
\begin{equation}
    \braket{\delta\theta_{x,{\rm FDMonly}}^2} = \frac{\lambda_{c}}{4\epsilon^{2}} \frac{\int dz\ \rho_{\rm NFW}^{2}(R)}{\Sigma_{\rm cr}^{2}} \label{eq:delta_theta2_fdmonly_no1}.
\end{equation}
Since the de Broglie wavelength is proportional to the inverse of the FDM mass $m$, we find the simple relation of $\braket{\delta\theta_{x,{\rm FDMonly}}^2} \propto 1/m$.
We can rewrite Eq.~\eqref{eq:delta_theta2_fdmonly_no1} as 
\begin{equation}
    %\frac{\braket{\delta\theta_{x,{\rm FDMonly}}^2}}{\theta_{\rm Ein}^{2}} = 
    \epsilon^{2}\braket{\delta\theta_{x,{\rm FDMonly}}^2}=
    \frac{\lambda_{\rm c}}{4r_{\rm h}} \kappa_{\rm FDM}^{2} \label{eq:delta_theta2_fdmonly_no2},
\end{equation}
where $r_{\rm h}$ is the effective radius in FDM halos introduced in Ref.~\cite{2022ApJ...925...61K}.
From Eq.~\eqref{eq:delta_theta2_fdmonly_no2}, the fluctuation of the macrocritical curve is found to be proportional to the convergence divided by the square root of the number of the clumps along the effective radius.

In the case where the baryon components distribute smoothly, the surface density power spectrum can be expressed with the same form as Eq.~\eqref{eq:Pkappa_fdmonly}, while the de Broglie wavelength is modified due to the additional baryon component.
% \begin{equation}
%     P_{\delta\kappa, {\rm FDMbaryon}} = \left(\frac{\Sigma_{\rm FDM}}{\Sigma_{\rm tot}}\right)^{2} P_{\delta\kappa, {\rm FDMonly}},
% \end{equation}
% where $\Sigma_{\rm tot}$ is the total (FDM + baryon) surface density.
The critical curve perturbation can be obtained as
\begin{equation}
%    \frac{\braket{\delta\theta_{x,{\rm FDMbaryon}}^2}}{\theta_{\rm Ein}^{2}} = 
     \epsilon^{2}\braket{\delta\theta_{x,{\rm FDMbaryon}}^2}=
    \frac{\lambda_{\rm c}}{4r_{\rm h}} \left(\frac{\kappa_{\rm FDM}}{\kappa_{\rm tot}}\right)^{2} \kappa_{\rm tot}^{2} \label{eq:delta_theta2_fdmbaryon}, 
\end{equation}
where $\kappa_{\rm tot} = \kappa_{\rm FDM} + \kappa_{\rm baryon}$ denotes the convergence of the total mass.
We can find that the smooth baryon profile reduces the fluctuation of the macrocritical curves.
% Note that the de Broglie wavelength in Eq.~\eqref{eq:delta_theta2_fdmbaryon} is not the same as that in Eq.~\eqref{eq:delta_theta2_fdmonly_no2} since the velocity is different due to the additional baryon component.

\begin{figure}[htbp]
    \centering
    \includegraphics[width=8cm]{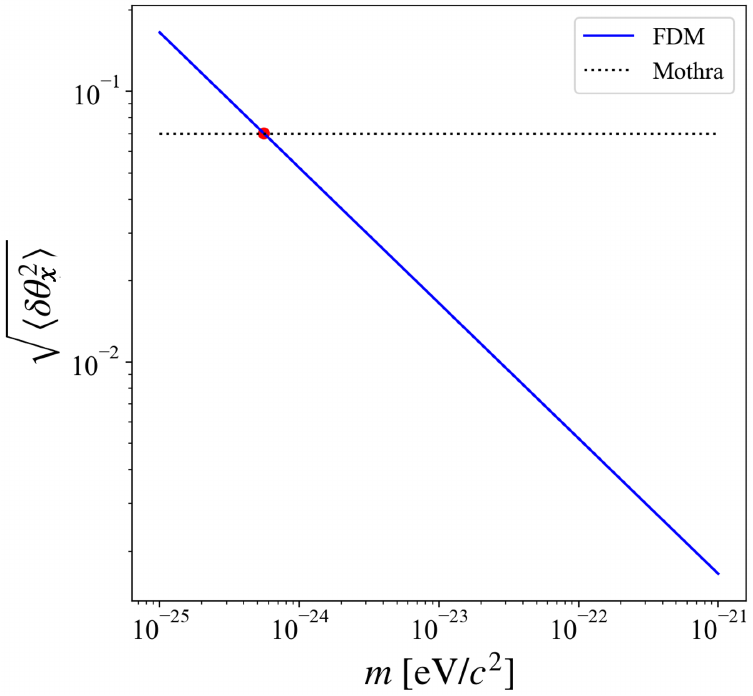}
    \caption{The relation between the critical curve perturbation and the FDM mass in the case of Mothra.
    %We set the halo mass $M_{\rm h} = 1.24 \times 10^{15}\ M_{\odot}$, the stellar mass $M_{\rm s} = 1.09 \times 10^{12}\ M_{\odot}$, the halo concentration $c_{h, \rm vir} = 7.59$, the effective radius in Hernquist profile $\theta_{\rm b} = 1.22"$, and Einstein radius $\theta_{\rm Ein} = 24.13"$.
    %Note that we use $\epsilon = 1/8["]$ instead of $\epsilon = 1/\theta_{\rm Ein}$ since the shape of the macrocritical curve is distorted. 
    The horizontal dotted line shows the fluctuation needed to explain the observed offset of Mothra.}
    \label{fig:fdm_mothra}    
\end{figure}

Using these relations, we calculate the critical curve perturbation in the specific case, Mothra. 
Figure~\ref{fig:fdm_mothra} shows the relation between the FDM mass and the fluctuation of the macrocritical curve. 
It is shown that the FDM mass of $mc^{2} \simeq 5.5\times 10^{-25}\ {\rm eV}$ is needed to explain the Mothra, which is significantly smaller than the typical FDM mass around $mc^{2} = 10^{-23}$--$10^{-21}\ {\rm eV}$.
We need a relatively small FDM mass to produce a sizable effect because, in galaxy clusters, the de Broglie wavelength is small due to large velocity dispersions and also, the averaging effect is larger due to the larger projection length along the line of sight. 
The preferred mass $mc^{2} \simeq 5.5\times 10^{-25}\ {\rm eV}$ is actually constrained by several works such as the analysis of the Lyman-$\alpha$ forest \citep{2017MNRAS.471.4606A,2017PhRvL.119c1302I,2019MNRAS.482.3227N,2021PhRvL.126g1302R}, which might indicate that the case where all DM is composed by FDM is ruled out. Although this case might not be valid, we could show an example of constraining DM properties based on the fluctuations of the main critical curve.
In order to provide an advanced statement for DM properties, we may need further discussions of hybrid dark matter models~(e.g., FDM+CDM), where we expect our formalism to be useful.
% Although this case might not be valid, we could show an example of constraining DM properties based on the fluctuations of the main critical curve.
% Further study that considers the case where the DM consists of several components ({\it e.g.,} FDM+CDM) is needed, where we expect our formalism to be useful.}

\section{Conclusion}\label{sec: conclude}
Astrometric perturbations of critical curves in strong lens systems serve as one of the most promising probes of small-scale substructures.
A smooth mass distribution creates a geometry that yields multiple images with symmetric configurations around critical curves with radii of curvature about the Einstein radius. 
Substructures introduce small-scale fluctuations on the macrocritical curves created by smooth mass distributions, which indicates
distortions or breaking the symmetry appearing in the macrolens model.
In this work, we have derived a general formula connecting fluctuations in the macrocritical curve with the fluctuations of the surface density due to substructures. This formula given in Eq.~\eqref{eq: dtheta2_dkappa2} allows us to analytically estimate the amplitude of the fluctuations from the surface density power spectrum of substructures. 
%In Sec.~\ref{sec: fluct_cc}, we have first reviewed the macrolens model. For simplicity, we have set coordinate systems in the image and source planes whose origin is a point on the critical curve and the caustic, respectively.
%Then, we introduced substructures in the macrolens model, yielding perturbations to a Jacobian matrix. 
%Since a critical curve is a line where the determinant of a Jacobian matrix becomes zero, the perturbations fluctuate the macrocritical curve. 
%We have analytically calculated the fluctuation by solving an equation for the critical curve.
%Our formula given in Eq.~\eqref{eq: dtheta2_dkappa2} between macrocritical curve fluctuations and fluctuations of surface density by substructures in Eq.~\eqref{eq: dtheta2_dkappa2}. Although there are previous works that have derived a formula between fluctuations of image positions and the surface density perturbations~\cite{2002ApJ...572...25D}, this is the first time the formula between the critical curve fluctuations and the surface density perturbations has been derived.
%In Sec.~\ref{sec: validity_cc}, 

We have explicitly checked the validity and accuracy of the formula in Eq.~\eqref{eq: dtheta2_dkappa2} using an open source code {\tt Glafic}.
Distributing subhalos near the macrocritical curve by the Poisson distribution, we have numerically computed critical curves with several models with different mass ranges of subhalos and calculated fluctuations of the macrocritical curves.
We have found that our formula is indeed valid and accurate as long as substructures are not dominated by a small number of massive structures. The numerical results are summarized in Table~\ref{tab: results}.

As a demonstration of our analytic formula, we have explored the possibility that an extremely magnified binary star recently reported in Ref.~\cite{2023arXiv230710363D} with JWST/NIRCam data, Mothra, can be explained by an unresolved magnified star whose position is offset due to the fluctuation of the macrocritical curve caused by substructures. We have found that CDM subhalos with masses ranging from $5\times 10^7~\Ms/h$ to $10^9~\Ms/h$ can well explain the anomalous position of Mothra.
On the other hand, we have found that the FDM with a very small mass of $\sim 10^{-24}$~eV is needed to explain Mothra. 

We expect that our analytic approach will be useful for studying fluctuations of macrocritical curve probed by highly magnified stars~\cite{2018ApJ...867...24D,2023arXiv230406064W} as well as by detailed mass modeling analysis of strong lensing systems~\cite{2020MNRAS.495.3192D,2023NatAs...7..736A,2023ApJ...954..197I,2023MNRAS.524L..84P}.

\hspace{3mm}
\acknowledgments
This work was supported by JSPS KAKENHI Grant No. JP22H01260, No. JP20H05856, No. JP22K21349, and No. JP22J21440.
% \HK{H.~K. thanks the hospitality of INAF OAS Bologna where part of this work was carried out.}

\appendix
\section{THE FOURIER TRANSFORMATION OF $\kappa$}\label{appendix: fourier_trans_kappa}
The two-dimensional Fourier transformation of the two-dimensional surface density fluctuation is written by
\eq{
\tilde{\kappa}_{\bm{k}}=\int d\bm{x} \kappa(\bm{x}) e^{\-i\bm{k}\cdot\bm{x}}.
}
\begin{widetext}
\eq{
\braket{\tilde{\kappa}_{\bm{k}}\tilde{\kappa}_{\bm{k}^\prime}}&= \int d\bm{x}\int d\bm{x}\prime~\braket{\kappa(\bm{x})\kappa(\bm{x}^\prime)}e^{-i\bm{k}\cdot\bm{x}}e^{-i\bm{k}^\prime\cdot\bm{x}^\prime}\\
&\ = \int d\bm{X}\int d\bm{X}^\prime~\frac{1}{\Sigma_{\mr{cr}}^2}\braket{\rho(\bm{X})\rho(\bm{X}^\prime)}e^{-i\bm{K}\cdot\bm{X}}e^{-i\bm{K}^\prime\cdot\bm{X}^\prime}|_{K_{Z}=K_{Z^\prime} = 0}\\
&\ = \frac{1}{\Sigma_{\mr{cr}}^2}\int d\bm{X}\int d\bm{X}^\prime~e^{-i\bm{K}\cdot\bm{X}}e^{-i\bm{K}^\prime\cdot\bm{X}^\prime}\left\langle\sum_{i} m_i^2 u(\bm{X}-\bm{X}_i|m)u(\bm{X}^\prime-\bm{X}_i|m)\right\rangle|_{K_{Z}=K_{Z^\prime} = 0}\\
&\ = \frac{1}{\Sigma_{\mr{cr}}^2}\int d\bm{X}\int d\bm{X}^\prime\int d\bm{X}^{\prime\prime}\int dm~\frac{dn(\bm{X}^{\prime\prime})}{dm} m^2 u(\bm{X}-\bm{X}^{\prime\prime}|m)u(\bm{X}^\prime-\bm{X}^{\prime\prime}|m)\\
&\ ~~~~~\times e^{-i\bm{K}\cdot(\bm{X}-\bm{X}^{\prime\prime})}e^{-i\bm{K}^\prime\cdot(\bm{X}^\prime-\bm{X}^{\prime\prime}))}e^{-i(\bm{K}+\bm{K}^\prime)\cdot\bm{X}^{\prime\prime}}|_{K_{Z}=K_{Z^\prime} = 0}\\
&\ = \frac{1}{\Sigma_{\mr{cr}}^2}\int d\bm{X}^{\prime\prime}~e^{-i(\bm{K}+\bm{K}^\prime)\cdot\bm{X}^{\prime\prime}}\int dm~\frac{dn(\bm{X}^{\prime\prime})}{dm} m^2 \tilde{u}_{\bm{K}}(m)\tilde{u}_{\bm{K}^\prime}(m)|_{K_{Z}=K_{Z^\prime} = 0}\\
&\ = \frac{(2\pi)^2}{\Sigma_{\mr{cr}}^2}\int dZ^{\prime\prime}\int dm~\frac{dn(Z^{\prime\prime})}{dm} m^2\tilde{u}_{\bm{k}}(m)\tilde{u}_{\bm{k}^\prime}(m) \delta^{\mr{2D}}(\bm{k}+\bm{k}^\prime)\\
&\ =  (2\pi)^2\delta^{\mr{2D}}(\bm{k}+\bm{k}^\prime)\frac{1}{\Sigma_{\mr{cr}}^2}\int dm~\frac{dn^{\mr{2D}}}{dm} m^2\tilde{u}_{\bm{k}}(m)\tilde{u}_{\bm{k}^\prime}(m), 
}
\end{widetext}
where $\bm{x}$ and $\bm{X}$ are two- and three-dimensional coordinates, respectively, $\bm{k}=(k_x,k_y)$, $\bm{K}=(k_x, k_y, K_z)$, and we assume that the spatial correlations among subhalos can be neglected. We also assume that the subhalo number density depends only on the coordinate along the line of sight, i.e. redshift. In this equation, we denote
\eq{\label{eq: uk_3d_to_uk_2d}
\tilde{u}_{\bm{K}}|_{K_{z}=0}=\tilde{u}_{\bm{k}}.
}

On the other hand, we can calculate $\braket{\tilde{\kappa}_{\bm{k}} \tilde{\kappa}_{\bm{k}^\prime}}$ directly like
\begin{widetext}

\eq{
\braket{\tilde{\kappa}_{\bm{k}} \tilde{\kappa}_{\bm{k}^\prime}}&= \int d\bm{x}\int d\bm{x}\prime~\braket{\kappa(\bm{x})\kappa(\bm{x}^\prime)}e^{-i\bm{k}\cdot\bm{x}}e^{-i\bm{k}^\prime\cdot\bm{x}^\prime}\\
&\ = \frac{1}{\Sigma_{\mr{cr}}^2}\int d\bm{x}\int d\bm{x}^\prime~\int d\bm{x}^{\prime\prime} \int dm \frac{dn^{\mr{2D}}}{dm}m^2 u^{\mr{2D}}(\bm{x}-\bm{x}^{\prime\prime}|m)u^{\mr{2D}}(\bm{x}^\prime-\bm{x}^{\prime\prime}|m)\\
&\ ~~~~~\times e^{-i\bm{k}\cdot(\bm{x}-\bm{x}^{\prime\prime})}e^{-i\bm{k}^\prime\cdot(\bm{x}^\prime-\bm{x}^{\prime\prime})}e^{-i(\bm{k}+\bm{k}^\prime)\cdot\bm{x}^{\prime\prime}}\\
&\ = (2\pi)^2\delta(\bm{k}+\bm{k}^{\prime})\frac{1}{\Sigma_{\mr{cr}}^2} \int dm \frac{dn^{\mr{2D}}}{dm}m^2\tilde{u}^{\mr{2D}}_{\bm{k}}(m)\tilde{u}^{\mr{2D}}_{\bm{k}^\prime}(m).
}
\end{widetext}
Then one can find that $u_{\bm{k}}$ represented in Eq.~\eqref{eq: uk_3d_to_uk_2d} is actually the two-dimensional Fourier transformation of the two-dimensional (normalized) surface density profile.

\bibliography{article}
\end{document}